\documentclass[aps,prl,twocolumn,groupedaddress,showpacs,floatfix]{revtex4}
\usepackage{graphicx}% Include figure files

\begin{document}

%\preprint{}

\title{
Critical Scattering and  Dynamical Scaling 
 in an Heisenberg Ferromagnet\\
  Neutron Spin Echo versus  Renormalization Group Theory
}

\author{Michel ALBA}
\email[]{michel.alba@cea.fr}
\affiliation{DSM-DRECAM, LLB CNRS-CEA, CEA-Saclay,
F-91191 Gif-sur-Yvette Cedex, France}

\author{St\'ephanie POUGET}
%\email[]{spouget@ill.fr}
\affiliation{CEA,  DSM-DRFMC, SP2M,  CENG, 17 rue des Martyrs,
F-38054 Grenoble Cedex 9, France}

\author{Peter FOUQUET}
%\email[]{fouquet@ill.fr}
\affiliation{Institut Laue-Langevin, 6 rue Jules Horowitz,
F-38042 Grenoble Cedex 9, France}

\author{Bela FARAGO}
%\email[]{farago@ill.fr}
\affiliation{Institut Laue-Langevin, 6 rue Jules Horowitz,
F-38042 Grenoble Cedex 9, France}

\author{Catherine PAPPAS}
%\email[]{pappas@hmi.de}
\affiliation{Hahn-Meitner Institut Berlin,
Glienicker Strasse 100, D-14019, Berlin, Germany}

\date{\today}

\begin{abstract}
High resolution Neutron Spin Echo  (NSE) spectroscopy was used to investigate  the dynamics of an 3D Heisenberg ferromagnet in the exchange-controlled regime over a broad range of temperatures and momentum transfer. These results allow for the first time an  extensive comparison between the experimental dynamical critical behavior and  the predictions of the Renormalization Group (RG) theory. The agreement is exhaustive and surprising as the  RG theory accounts not only for the critical relaxation but also for the shape crossover towards an exponential diffusive relaxation when moving from the critical to the hydrodynamic regime above $T_C$.
\end{abstract}

%\keywords{}

\maketitle


%%\section{Critical~dynamics of the 3D Heisenberg ferromagnet}
Critical phenomena 
are ubiquitous in condensed matter physics\cite{ma75} and their theories are now used to  understand   the behavior of many complex systems, 
like sandpiles \cite{Aeg04} or  tropical rains \cite{Peters06}.  
Phase transitions are dynamical processes in essence, specially  the continuous phase transitions. These  are characterized by diverging fluctuations at the critical temperature  $T_{C}$,
with a  critical slowing-down and a scale-free behavior: power law spatial and temporal evolutions of the order-parameter fluctuations.
In the critical region, the dynamics is governed by a single dimensionless variable,
according to the dynamical scaling hypothesis \cite{ferrell67}.
Ferromagnetism serves as a paradigm for continuous phase transitions
 and as such a lot of theoretical and experimental effort \cite{FREY94,fol05}
has been devoted to understand and predict in details its static and dynamic critical behavior.

\begin{figure}[!ht]
 \includegraphics[width=0.87 \columnwidth ]{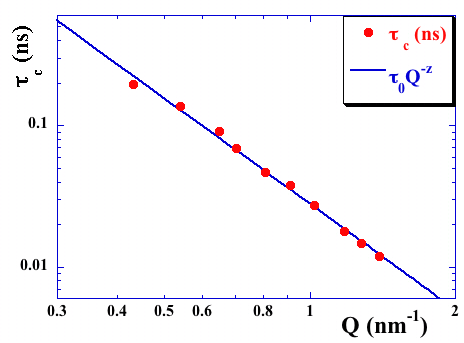}
  \caption{\label{tau_fit_final}
Critical slowing down of  the characteristic relaxation time $\tau_{c}$ of the
magnetic fluctuations in  $CdCr_{2}S_{4}$. The solid line is a  power law fit,
$\tau_{c} (Q,T_{C}) =\tau_{0} Q^{-z}$  with $z=2.47 \pm 0.02$  and $\tau_{0} = 0.02813~ns/nm^{-z} $. }
 \end{figure}

Critical dynamics is a much more complex problem than the static
critical phenomena as it depends not only on the static universality class
 but also on conservation laws and on the equations  of motion of the
 system  \cite{Hoh77}.  For example,
the 3D Heisenberg ferromagnet displays  two different behaviors,
depending on the relevant interactions, exchange or dipolar.
The balance between exchange or dipolar interactions is quantified
by the dipolar wave-vector $Q_{D}$
($Q_{D}^{2} = \mu_{0}k_{B} \frac{(g\mu_{B})^{2}}{v_{0} J_{2}}$,
$J_{2}$ being the second moment of the exchange interactions). 
Exchange  controls the dynamical behavior  for $Q > Q_{D}$
 and  the order-parameter (i.e. the magnetization) is conserved  (model J of Reference (\cite{Hoh77})).
The critical dynamics at $T_{C}$ 
 is  governed  by a dynamical exponent  z=$\frac{D+2-\eta}{2}$,  where  D is
 the space dimension and $\eta $ the Fisher exponent. For 3-D Heisenberg systems,
  $z \approx 5/2$ (figure \ref{tau_fit_final}). 
 Dipolar interactions, on the other hand, control the dynamics at  $Q<Q_{D}$. These interactions do not conserve the order parameter and at  $T_{C}$ the  dynamical exponent crosses over from z=5/2 (exchange) to z=2 (dipolar) at $Q \approx Q_{D}/10$. 
 
 In the critical region above $T_{C}$ and in the sole presence of exchange interactions, the relevant  scaling variable is $x= \kappa/Q$,  with  $\kappa $ the inverse magnetic correlation length.  In this region the magnetic fluctuations  decay faster than an exponential   with a shape cross-over   to an exponential relaxation when going from the critical to the hydrodynamic regime  for  $x\gg 1$.
This simple scaling hypothesis  breaks down  in the presence of dipolar interactions.   At $Q<Q_{D}$, the relaxation time depends on two reduced variables: $x= \kappa/Q$ and $y=Q/Q_{D}$  \cite{Mez82} whereas the relaxation recovers a simple exponential decay. The  boundary conditions for the exchange regime are therefore fixed by the theory: $Q_{D}<Q<\kappa$, which implies that the critical non-exponential relaxation can only be observed in a precise temperature and Q range and for ferromagnetic systems with low enough values of  $Q_{D}$. 

%% \subsection{The Renormalization~Group~theory} 
Independent renormalization Group (RG)  \cite {Iro89} and Mode Coupling
(MC)\cite{FREY94} calculations  predict similar shapes for the spectral function at  the exchange controlled regime at  $Q>Q_{D}$.  These predictions, however, were never tested thoroughly. One reason ist that  in neutron scattering experiments  the finite resolution blurs
out the magnetic relaxation spectra. Triple-axis neutron experiments
(see \cite{Boni93} and references therein) focussed on constant-energy  Q-scans which  displayed a broad maximum the position of which was compared with the theoretical calculations.
In this letter, we present the first detailed analysis of the shape of the relaxation function in a  model 3D Heisenberg ferromagnet by taking advantage  of the  recent progress in the field of high resolution Neutron Spin-Echo (NSE) spectroscopy.

The RG theory expresses  the spectral shape function $F(Q,\omega)$  as the
real part of a complex function  of the 
scaling variable $\kappa/Q$ and the scaled frequency $\omega/\omega_{c}$ ,  with the help of an $\epsilon=6-D$ expansion \cite{Iro89} and three parameters  $k$, $a$  (for the critical behavior) and $b$
 (for the cross-over to the hydrodynamic regime) :
\begin{equation}
F(Q,\omega) = \frac{2}{\omega_{c}} 
 \Re \left[  \left(  
\left(
Z(x)\Pi_{1}\left(x,iu\right)
\right) ^{-1} 
  -i \frac{\omega}{\omega_{c}}  \right) ^{-1}
\right]
\label{RG_funkt}
\end{equation}
The function $\Pi_{1}(x,iu)$ is the self-energy of the
dynamic susceptibility , approximated by :
\begin{equation}
\Pi_{1}\left( x,iu\right) = \left ( \left ( 1+bx^{2}
\right)^{2-\frac{ \epsilon}{4}} -i a u \right)^{\frac{\epsilon}{8-\epsilon}}
\label{Epsexp}
\end{equation}
(where $ u=\omega /\omega_{c} \times Z(x)/X(x)$ for clarity).
  $Z(x)$ is
essentially a  scaling function, approximated by:
\begin{equation}
 Z(x) = \frac {(1+bx^{2}) ^{\frac{-\epsilon}{4}}} { 1-k \times Arctan \left(
 a \frac {1+x^{2}}{(1+bx^{2})^{2}} \right) }
\label{ScaleZ}
\end{equation}
 and $X(x)$ is  the $\chi(Q)$ susceptibility  modelled by a  Lorentzian $1/(1+x^{2})$ 
  in the Ornstein-Zernicke approximation.
  The scaling function describing the crossover with $x$   from the critical behavior  to the diffusive one in  the hydrodynamic regime is given by :
 
  \begin{equation}
\Omega (x) = \frac{\omega_{c}(Q,T)} {\omega_{c}(Q,T_{c})} = \frac{\tau_{c}(Q,T_{c})} {\tau_{c}(Q,T)} = \frac{1}{X(x)}  \frac{Z(x)}{Z( 0 )}
 \label{ScalingF}
\end{equation}

  For a better comparison with the NSE experimental results, a analytical form of the intermediate
  Fourier transform of the spectral function \cite {folk93}  has been proposed:
 
 \begin{equation}
I_{RG}(Q,t )  = e^{-a (x) \frac{t}{\tau_{c}} }   \Big( Cos \big( b(x)  \frac{t}{\tau_{c}}  \big) + Sin\big( b(x)  \frac{t}{\tau_{c}}  \big)  \Big)
\label{TheoInter}
\end{equation}
with
 \begin{equation}
a(x)  = \frac{3.23+3.26 x+4.195 x^{2}} {1+0.28 x^{1/2}+3.84 x^{3/2}} \  ,\ 
 b(x) =  \frac{2.93} {1+2.1 x} 
 \label{coefs}
\end{equation}
and $\tau_{c}(Q,T) \propto  \Omega(x) \tau_{0}Q^{-z} $. We will use this convenient parametrization of the RG theory to reduce our NSE experimental data, at the critical temperature $T_{C}$ ($\Omega(x)=1$) and above ($\Omega(x)\neq1$).

%% \subsection{Structure and magnetic interactions of The Chromium thiospinel}
 $CdCr_{2}S_{4}$ is a  normal spinel, with all  Cr$^{3+}$ ions
  having a magnetic moment  of
 spin only type (S=$ 3\over 2$) with  a very low magnetic anisotropy.
$CdCr_{2}S_{4}$ is a semiconductor and undergoes a ferromagnetic transition 
 below $T_{C}=84.8~K$. The magnetic interactions are positive between 
the six nearest neighbors (${J_{1} / k_{B}}= 13.25~K$), negligible between the six next nearest neighbors and negative (${J_{3} / k_{B}}= -0.915~K$)  between the twelve third nearest 
neighbors. These interactions lead to a very small dipolar wave-vector of 
$Q_{D} = 0.51~nm^{-1}$,
 making this coumpond an ideal candidate for the study of the exchange governed critical dynamics.
Our comprehensive investigation of the static critical properties
(\cite{Pou95b} and references therein),
 determined the critical exponents 
 $ \beta = 0.33 \pm 0.03$, $\nu = 0.70 \pm 0.03$ , $\eta = 0.065 \pm 0.030$, 
 and $\gamma = 1.40 \pm 0.04$,
which are very close to the theoretical values for the 3D
Heisenberg model \cite{Guil89}. Therefore, 
$CdCr_{2}S_{4}$ is a textbook example of
the 3D Heisenberg ferromagnetic class.
%%: $\nu_{3D-H} = 0.710$ and
%%$\gamma_{3D-H} = 1.390$. We were also able to estimate the Fischer
%%exponent value, $\eta = 0.065 \pm 0.030$, compatible
%%with the theoretical prediction ($\eta_{3D-H} = 0.040$).

 %\section{Neutron~scattering}
  \begin{figure}[!ht]
 \includegraphics[width= 0.87\columnwidth ]{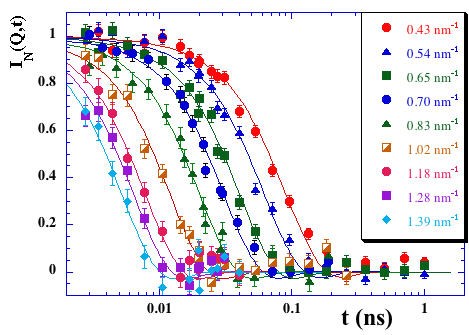}%
 \label{crit_slow_final} 
 \caption{
Magnetic critical scattering at $T_{C}$ in $CdCr_{2}S_{4}$ , measured on the NSE spectrometer IN11 (ILL). Lines are adjustments of the RG theory of  $I(Q,t)$  (see equation \ref{TheoInter}) }
 \end{figure}
The  measurements were performed on the 
neutron spin-echo (NSE) spectrometer  \textbf{IN11} 
at the Institut Laue Langevin(Grenoble, France). 
The polycrystalline samples were enclosed in a sealed aluminium can
and placed in a standard Orange cryostat, where the temperature was
controlled with a precision of $\pm 0.002~K$ in the $1.5-300~K$ range.
The transition temperature $T_{C}$ was determined from  the maximum of 
the diffused intensity at small Q in unpolarized mode,
with correction of the offset \cite{rit72} due to the Fisher exponent $\eta$ 
and the finite  Q value. 
The NSE spectra were recorded with a $0.45~nm$ neutron wavelength in the standard paramagnetic setup with a 2D multidetector whose pixels were divided in three constant-Q stripes. Separate measurements of  the direct beam polarization made sure  that any spurious  depolarization effects  from the sample were avoided. 
 The NSE spectra were measured for $1.001 ~T_{C} \le T \le 1.19 ~ T_{C}$ and $0.43 \le Q \le 1.39 ~nm^{-1}$,
 the scaling variable $x=\kappa/Q$ covering the range  $0.03-1.6$.
The NSE resolution was 
measured on a similar spin glass sample at a sufficiently low temperature for the
spin dynamics to  be  completely frozen and static on the probed timescale. 
All dynamic curves were divided by the corresponding 
 resolution function and  normalized by the instantaneous magnetic  response
determined as part of the NSE measurements by x-y-z polarization analysis.
This treatment gives the normalized intermediate scattering function  $I_{N}(Q,t)$
that can be compared with  the theoretical dynamical shape function  $I_{RG}(Q,t)$ (equation \ref{TheoInter}).

%% \section {critical dynamics at TC}
\begin{figure}[!ht]
 \includegraphics[width= 0.88\columnwidth ]{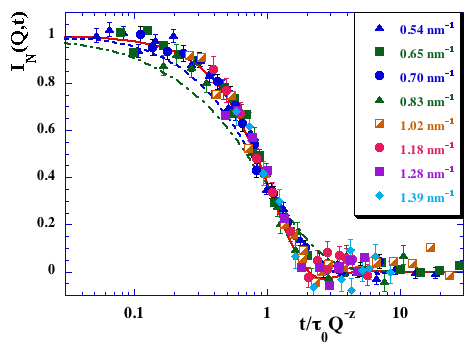}%
 \caption{\label{scalingpur04_expfinal}
Dynamical scaling of the magnetic critical scattering at $T_{C}$ in $CdCr_{2}S_{4}$ for $Q \geq Q_{D}$. The  red straight line displays the original RG model  (equation \ref{TheoInter}), the blue dash line a modified RG model\cite{alba01}  and the  green dash-dotted line a simple exponential decay of the magnetic correlations.}
 \end{figure}
  At $T_{C}$, the adjustments of the  $I_{RG}(Q,T)$ parametrization 
 to the NSE curves  led to the power law decay (figure \ref{tau_fit_final})
 of  the characteristic relaxation time
$\tau_{c} =\tau_{0} Q^{-z}$  with $z=2.47$ and $\tau_{0} = 0.02813~ns/nm^{-z} $.
Figure \ref{scalingpur04_expfinal} shows that all
spectra   collapse in a master curve when plotted against 
 the scaling variable $t/ \tau_{0} Q^{-z}$.
The decay  of the relaxation is faster than an exponential and is surprisingly well accounted for by the   original RG calculation, even though it is only a first order $\epsilon$-expansion. 
A modified RG model\cite{alba01}  is less convincing.
We will comment on  these modified RG  versions later in the discussion.
%% \section {dynamic scaling  above TC}
%%$x=\kappa_{0}((T-T_{C})/T)^{\nu}/Q$ 
 \begin{figure}[!ht]
 \includegraphics[width= 0.88\columnwidth ]{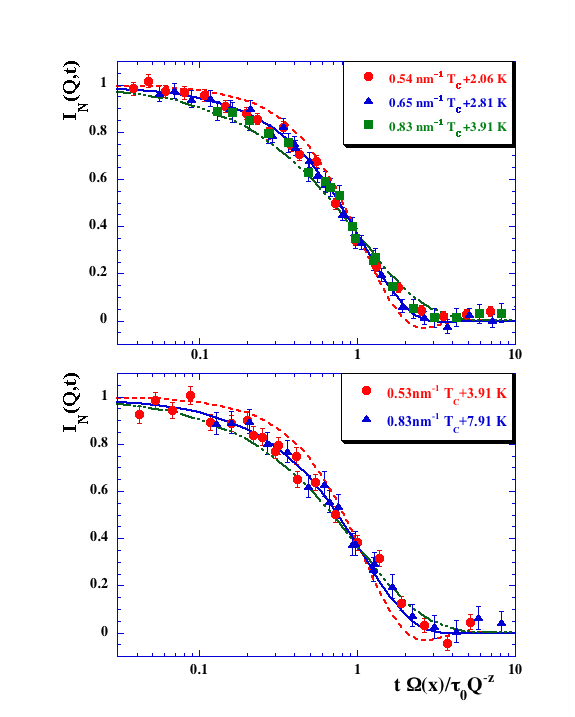}
\caption{\label{scaling_layout}
Crossover  of the shape function  from the critical towards an exponential decay
 as a function of the reduced variable $x= \kappa/Q$ at  $x=0.49$  (upper figure)
 and at $x=0.77$ (the lower figure). The blue solid lines display the RG theory at the corresponding 
  $x$ value, the red dash lines the critical shape and 
  the green dash-dotted lines and exponential decay. }
 \end{figure}

As mentioned previously, we went beyond the trivial verification of the dynamical scaling
at $T_{C}$,  where $\kappa /Q  = 0$, and measured the magnetic relaxation above $T_{C}$ at finite values of $x=\kappa/Q$, where the function departs from the critical shape and evolves
towards an exponential. The spectra were measured at constant $x=\kappa/Q$  and scaled with the reduced  time $t/\tau(Q,x) =t \Omega(x)/ \tau_{0} Q^{-z} $. 
The experimental results are displayed in figure \ref{scaling_layout} for  $x=0.48$ and $x=0.77$
respectively, with $\Omega(0.48)=0.704$ and  $\Omega(0.77)=0.590$. For each  $x$,  the magnetic relaxation spectra collapse on a single master curve, the shape of which evolves with increasing $x$ towards the exponential.
 These master curves are well described by  equation  \ref{TheoInter}
with the measured critical exponents $\nu = 0.7\pm0.03 $,  $z= 2.47 \pm0.02 $ ,  and critical parameters: transition temperature  $T_{C}= 84.84 \pm 0.004~K$ and amplitudes $\kappa_{0}= 3.61~nm^{-1}$ and $\tau_{0} = 0.02813~ns/nm^{-z} $.

%% \section {discussion}
\begin{figure}[!ht]
 \includegraphics[width= 0.87\columnwidth ]{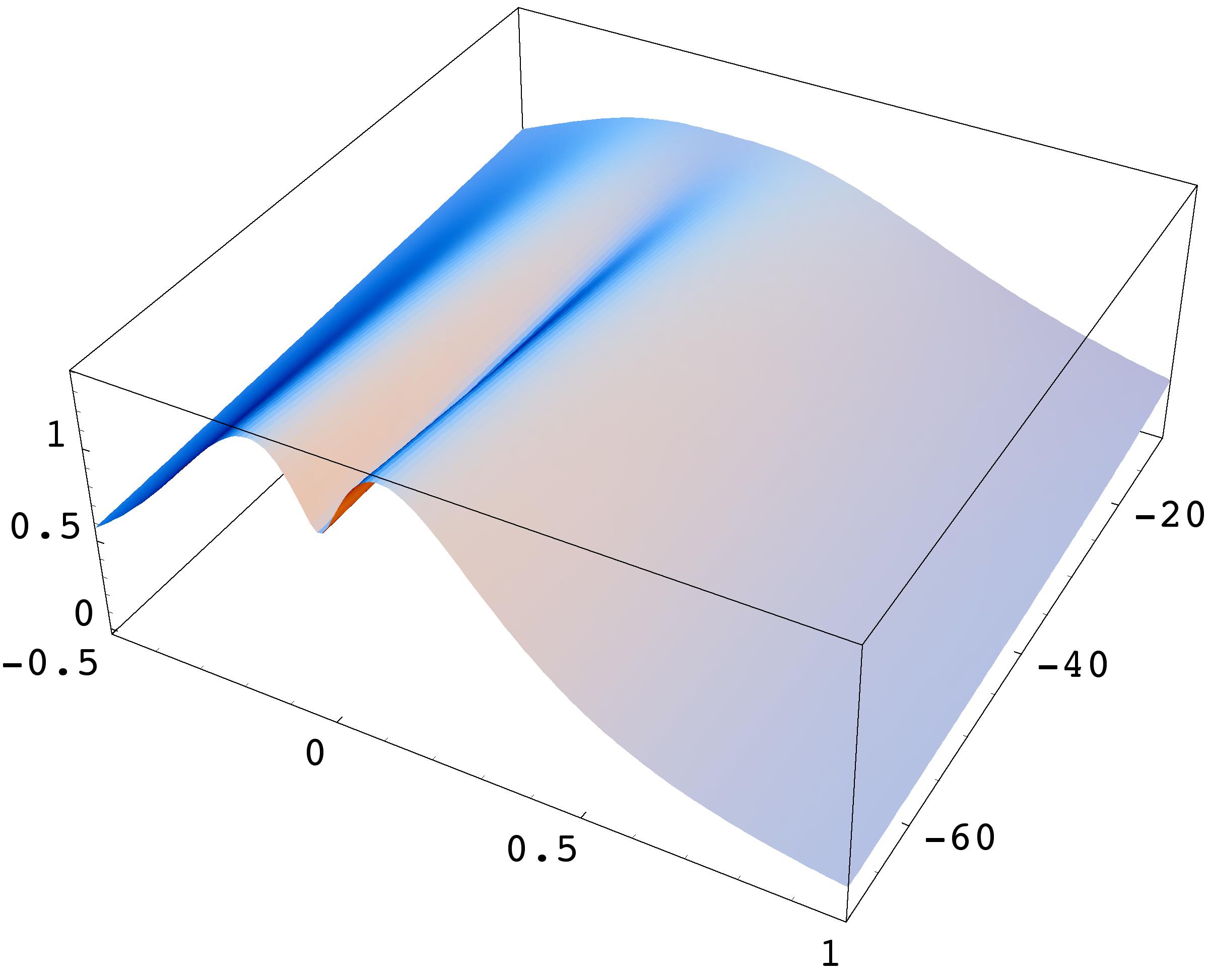}%
 \caption{\label{crit_m20}
 Critical scattering $F(Q,\omega)$ of equation \ref{RG_funkt}  for $k=0.51$ and  $-70<a<-10$,
 plotted for $-0.5<\omega/\omega_{c}<1$ and 
 showing an overdamped maximum for $a<-10.75$.}
 \end{figure}
  
 This exhaustive agreement between experiment and theory is new and in some sense surprising. Previous attempts to test the critical dynamic   theory with neutron triple axis spectroscopy \cite{Boni93, alba01}  adjusted the parameters $a$, $k$ and $b$ to the measured scaling function $\Omega(x)$  over an extended $x$ range, leading to large negative values for the parameter  $a$. These  induce
a broad overdamped maximum in $F(Q,\omega)$ (figure \ref{crit_m20}) which is definitely not compatible with experiments and our high resolution NSE experiments (figure \ref{scalingpur04_expfinal}) when Fourier transformed in time.
Obviously, these parameter adjustments are questionable as they are done in a much too large $x$ range. In fact, only pure Iron follows the theoretical curve over such an extended range \cite{Mez82}.
A self consistent reparametrisation would imply a one-step iteration to extract new $\omega_{c} (Q,T)$ values with the reparametrized  $F(Q,\omega)$  and get a new scaling function $\Omega(x)$  that
would converge to a stable shape and set of parameters. Moreover, at large $x$ values, only  the $b$ parameter  might be adjusted \cite {Iro89} whereas the  $a$ parameter should be kept  between   0.46 and 0.99 \cite{kala90}  or at least  between -10.75 and 1 to avoid a broad maximum in $F(Q,\omega)$. The discrepancy between the neutron triple-axis results and the original RG predictions
could therefore be used for a self consistent reparametrisation but even in this case the conclusions would be limited by the convolution of the complex function of equation \ref{RG_funkt} with the  triple axis resolution function.

% Conclusions ...
Neutron Spin Echo is a unique technique for the detailed analysis of relaxation shape functions, especially magnetic ones. In NSE resolution corrections reduce to a simple division. Moreover, the paramagnetic NSE setup discards all nuclear parts of the scattering and extracts directly the critical intermediate functions, which are as exact as possible. This simple and straightforward data reduction  allowed us to check the dynamical scaling on the shape of the relaxation over an extended temperature and Q range.
 Our NSE study shows that the RG theory gives an accurate quantitative description of the critical scattering and dynamical scaling  not only at  $x=0$ but also at the crossover to the hydrodynamic regime at $x \leq 1$.  At larger $x$ values, however, the theory may not be applicable without reparametrization  either due to its approximations or to deviations of the real systems from the 3D Heisenberg model. 

 Similar studies were recently performed  on  Ising systems  showing an exponential decay on the microsecond or second timescales using Photon Correlation Spectroscopy 
with  either X-rays \cite{mocuta05} or visible light \cite{Iwan06}. 
 However, the most detailed dynamic studies were performed by NSE \cite{Lecler05} on the nanosecond timescale.
 In spin glasses, dynamical scaling is more a question   of phase space  rather than reciprocal space,  with streched exponential  and power law decay \cite{pappas03} over an extended time range.  The agreement between the experimental results and the theoretical predictions presented in this letter goes beyond these findings. It is  the first clear experimental validation  of  all the implications of the dynamical scaling hypothesis  not only  for the characteristic relaxation times but also for the very shape of the relaxation function, in the nanosecond and nanometer range.
\begin{acknowledgments}
%We would like to thank Peter B\"{o}ni  for useful discussions.
\end{acknowledgments}


\end{document}